\documentclass[12pt,draftclsnofoot,onecolumn]{IEEEtran}
\usepackage{amssymb}
\usepackage[cmex10]{amsmath}
\usepackage{stfloats}
\usepackage{graphicx}
\usepackage{subfigure}
\usepackage{tabularx}
\usepackage{epsfig,epsf,color,balance,cite}
\usepackage{verbatim}

\newtheorem{theorem}{Theorem}
\newtheorem{lemma}{Lemma}
\usepackage{algorithm}
\usepackage{algorithmic}
\usepackage{amsmath}

\usepackage{color}
\definecolor{myc1}{rgb}{0,0,0}
\definecolor{myc2}{rgb}{0,0,0}
\begin{document}

\title{Efficient Resource Allocation for Mobile-Edge Computing Networks with NOMA:  Completion Time and Energy Minimization}

\author{
\IEEEauthorblockN{Zhaohui Yang,
                  Cunhua Pan,
                  Jiancao Hou,
                  and Mohammad Shikh-Bahaei
                  }
\thanks{Z. Yang, J. Hou and M. Shikh-Bahaei are with the Centre for Telecommunications Research, Department of Informatics, King’s college London, London WC2B 4GB,  U.K., (e-mail: \{yang.zhaohui, jiancao.hou, m.sbahaei\}@kcl.ac.uk).
}
 \thanks{C. Pan is
 with the School of Electronic Engineering and Computer Science, Queen Mary, University of London, London E1 4NS, U.K., (e-mail:
 c.pan@qmul.ac.uk).}
}

\maketitle

\begin{abstract}

This paper investigates an uplink non-orthogonal multiple access (NOMA)-based mobile-edge computing (MEC) network. Our objective is to minimize a linear combination of the completion time of all users' tasks and the total energy consumption of all users including transmission energy and local computation energy subject to computation latency, uploading data rate, time sharing and edge cloud capacity constraints. This work can significantly improve the energy efficiency and end-to-end delay of the applications in future wireless networks. For the general minimization problem, it is first transformed into an equivalent form. Then, an iterative algorithm is accordingly proposed, where closed-form solution is obtained in each step. For the special case with only minimizing the completion time, we propose a bisection-based algorithm to obtain the optimal solution. Also for the special case with infinite cloud capacity, we show that the original minimization problem can be transformed into an equivalent convex one. Numerical results show the superiority of the proposed algorithms compared with conventional algorithms in terms of completion time and energy consumption.
\end{abstract}
\IEEEpeerreviewmaketitle
 \newpage
\section{Introduction}
Mobile-edge computing (MEC) has been deemed as a promising technology for future communications due to the fact that it can improve the computation capacity of users in applications, such as, augmented reality (AR) \cite{8016573,saad2019vision}.
With MEC, users can offload the tasks to the MEC servers located at the edge of the network.
Since the MEC servers can be deployed near to the users, a network with MEC can provide users with low latency and low energy consumption~\cite{7906521}.

The basic idea of MEC is to utilize the powerful computing facilities within the radio access network, such as the MEC server integrated into the base station (BS).
Users can benefit from offloading the computationally intensive tasks to the MEC server \cite{pham2019survey}.
There are two operation modes for MEC, i.e., partial and binary computation offloading.
In partial computation offloading, the computation tasks can be divided into two parts, where one part is locally executed and the other is offloaded to the MEC server \cite{8006982,7762913,8254208,7929399,8240666,8168252,7842016}.
In binary computation offloading, the computation tasks are either locally executed or completely offloaded to the MEC server \cite{8314696,6574874,8330749,8305608}.

Due to limited radio resources of the wireless links and computation resources at the edge cloud, it is of importance to investigate resource allocation for MEC networks.
Two common resource allocation problems have been considered for MEC: completion time minimization \cite{8006982,8314696} and total energy minimization \cite{7762913,6574874,8267072,8330749,8305608,8274943}.
Two joint resource allcoation algorithms were developed in \cite{8006982} for minimizing the tasks' completion time including the time for data transmission and computing in time division multiple access (TDMA) and frequency division multiple access (FDMA) schemes.
To minimize the total energy of all users, joint time allocation and power control was optimized for a TDMA-based MEC network in \cite{7762913}.

Recently, non-orthogonal multiple access (NOMA) has been recognized as a potential technology for the next generation wireless mobile communication networks to tackle the explosive growth of data traffic \cite{vaezi2018multiple,vaezi2019interplay,Zhiguo2017Survey,7890454,7263349,wang2018user,yang2018cooperative,8543183,8403960,8647935,Yang2018Power}.
Due to superposition coding at the transmitter and successive interference cancelation (SIC) at the receiver, NOMA can achieve higher spectral efficiency than conventional orthogonal multiple access (OMA), such as TDMA and FDMA \cite{Yang2017Joint}.
Many previous contributions \cite{7906521,8006982,8254208,7762913,7929399,8240666,8168252,7842016} only considered OMA.
Motivated by the benefits of NOMA over OMA, a NOMA-based MEC network was investigated in \cite{8269088}, where users simultaneously offload their computation tasks to the BS and the BS uses SIC for information decoding.
Besides, both NOMA uplink and downlink transmissions were applied to MEC \cite{zhi2018ImNOMAMEC}, where analytic results were developed to show that  the latency and energy consumption can be reduced by applying NOMA-based MEC offloading.
Completion time minimization and energy minimization were respectively optimized in \cite{Ding2018Delay} and \cite{Ding2018Joint} for NOMA-based MEC networks with different computation deadline requirements for different users.
However, the authors in  \cite{8269088,zhi2018ImNOMAMEC,Ding2018Delay,Ding2018Joint} only considered one group of users forming NOMA and ignored the time allocation among different groups of users forming NOMA.
Since each resource is recommended to be multiplexed by a small number of users (for example, two users) due to decoding complexity and error propagation \cite{6464495}, it is of importance to investigate  resource allocation among different groups of users forming NOMA.

In this paper, we investigate the resource allocation for an uplink NOMA-based MEC network. To our best knowledge, this is the first work that investigates the resource allocation for NOMA-based MEC network by considering multiple groups with multiple users in each group.
The main contributions of this paper are summarized as follows:
\begin{enumerate}
  \item  A linear combination of the completion time and the total energy consumption is minimized for an uplink NOMA-based MEC network.   Different from \cite{8269088,zhi2018ImNOMAMEC,Ding2018Delay,Ding2018Joint}, time allocation for different groups is investigated in this paper, where multiple users are clustered in each group to perform NOMA.
 Different from our conference paper \cite{yang2018gc}, this paper considers multiple users in different groups to share the radio resource in NOMA as in \cite{8403960,8647935}.
 \item To solve the minimization problem, an iterative algorithm with low complexity is proposed, where the closed-form solution is obtained in each step.
  \item  For the special case with only minimizing the completion time, a bisection-based algorithm is accordingly proposed to obtain the optimal solution, which requires to solve the feasibility problem in each iteration.  For the special case with infinite edge cloud capacity, the original minimization problem is shown to be equivalent to a convex one. It is also proved that transmitting with maximal time is always energy efficient.
\end{enumerate}

The rest of the paper is organized as follows.
In Section $\text{\uppercase\expandafter{\romannumeral2}}$, we introduce the system model and formulate the optimization problem.
Section $\text{\uppercase\expandafter{\romannumeral 3}}$ solves the energy efficient resource allocation problem.
Numerical results are shown in Section $\text{\uppercase\expandafter{\romannumeral 4}}$
and conclusions are finally drawn in Section~$\text{\uppercase\expandafter{\romannumeral 5}}$.

\section{System Model and Problem Formulation}
\begin{figure}
\centering
\includegraphics[width=3.5in]{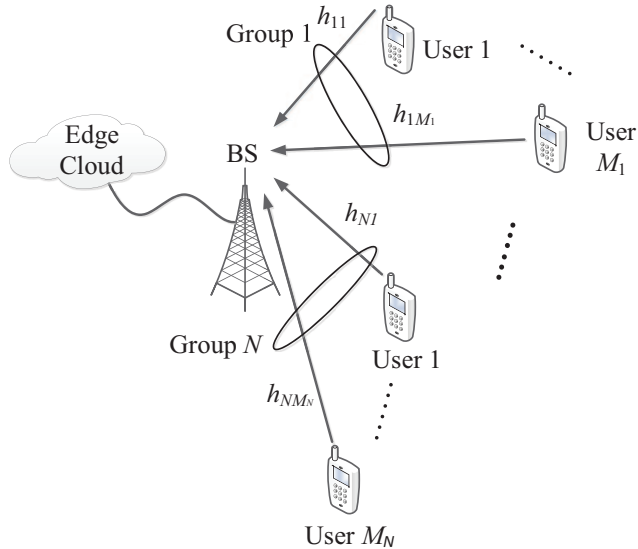}
\vspace{-1.5em}
\caption{Multi-user MEC network with NOMA.}\label{sys1fig1}
\vspace{-1.75em}
\end{figure}

Consider a NOMA-enabled MEC network with $M$ users and one BS that is the gateway of an edge cloud, as shown in Fig.~\ref{sys1fig1}.
All $M$ users are classified into $N$ groups with $M_i$ users in group $i$\footnote{In this paper, we assume that the user grouping is given, which can be obtained by the matching theory \cite{Fang2016EE} or according to the order of channel gains \cite{7273963}. {\color{myc2}{Note that it is also possible to apply any grouping schemes and the system performance is highly dependent on the grouping method. Since the main novelty of this paper is the optimization of data offloading, time allocation and power control, the optimization of user grouping is beyond the scope of this paper.}}}, i.e., $M=\sum_{i}^N M_i$.
Let $\mathcal N=\{1, 2, \cdots, N\}$ denote the set of all groups and $\mathcal J_i=\{1, 2,\cdots,M_i\}$ denote $M_i$ users in group $i$.
The users in each group simultaneously transmit data to the BS at the same frequency by using NOMA \cite{6464495}.
We consider TDMA scheme for users in different groups. 

The BS schedules the users to partially or completely offload tasks.
The users with partial or complete offloading respectively offload a fraction of or all input data to the BS, while the users with partial or no offloading respectively compute a fraction of or all input data using local central processing unit (CPU).
ther are some examples about partial computation offloading, such as  the terrorist detection and  find-missing-children in \cite{7786106}.
In the terrorist detection or find-missing-children application, the users have many photos to be computed (i.e., the face matching computation), which can take a lot of time if all the photos are computed locally.
In this case, each user can send part of the photos to the BS and the BS
 performs face matching to compare the
terrorist's or missing-child's photo with the photos taken by users.
After finishing the computation at the BS, the BS transmits the computation results (whether the terrorist or missing-child is found or not) to the users.
Due to the small sizes of computation results, the time of  downloading from the BS is negligible compared to the time of  offloading and computing \cite{7842016}.

The BS is assumed to have the perfect information of uplink channels, local computation capabilities, power limits and input data sizes of all users  \cite{7762913}.
All channels are assumed to be frequently flat.
Using this information, the BS determines the offloaded data, time sharing factor,  time allocation,  computation capacity allocation, and transmission power of all users.
\vspace{-.75em}
\subsection{Local Computing Model}
\vspace{-.25em}

The local computing model is described as follows.
Denote $R_{ij}$ (bits) as the total input data of user $j$ in group $i$.
Since only $d_{ij}$ bits are offloaded to the BS, the remaining $R_{ij}-d_{ij}$ bits are needed to be computed locally at user $j$ in group $i$.
Based on the local computing model in  \cite{7842016}, the total energy consumption for local computation at user $j$ in group $i$ is
\vspace{-0.5em}
\begin{equation}\label{sys1eq1}\vspace{-0.5em}
E_{ij}^{\text{Loc}} =  C_{ij}Q_{ij}(R_{ij}-d_{ij}), \quad \forall i\in\mathcal N, j\in\mathcal J_i,
\end{equation}
where $C_{ij}$ (cycles/bit) is the number of CPU cycles required for computing 1-bit input data at user $j$ in group $i$,
and $Q_{ij}$ (J/cycle) stands for the energy consumption per cycle for local computing at this user.

Let $F_{ij}$ denote the computation capacity of user $j$ in group $i$, which is measured by the number of CPU cycles per second.
The processing time of the local job at user $j$ in group $i$ is
\vspace{-0.5em}
\begin{equation}\label{sys1eq1_2}\vspace{-0.5em}
T_{ij}^{\text{Loc}} =\frac{C_{ij}(R_{ij}-d_{ij})  }{F_{ij}} , \quad \forall i\in\mathcal N, j\in\mathcal J_i.
\end{equation}
\vspace{-1em}
\subsection{Transmission Scheme}
\vspace{-0.5em}
Denote the bandwidth of the network by $B$, and the power spectral density of the additive white Gaussian noise by $\sigma^2$.
Let $h_{ij}$ denote the channel gain between user $j$ in group $i$ and the BS.
Without loss of generality, the uplink channels between users in group $i$ and the BS are sorted as $h_{i1} \geq  h_{i2}\geq\cdots\geq h_{iM_i}$, $\forall i\in \mathcal N$.
Denote $p_{ij}$ as the transmission power of user $j$ in group $i$.

\begin{figure}
\centering
\includegraphics[width=6in]{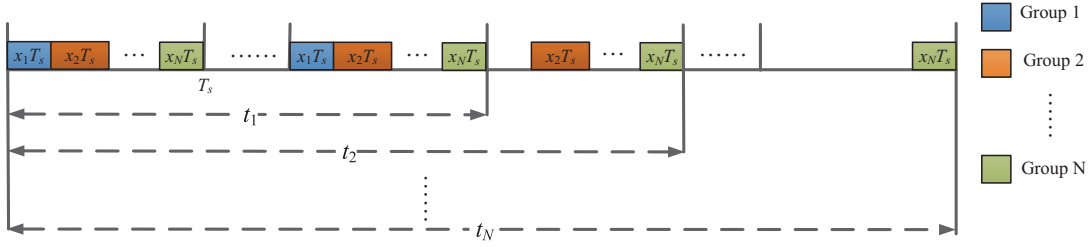}
\vspace{-1.5em}
\caption{{\color{myc1}{A special example of the TDMA scheme with $t_1<t_2<\cdots<t_N$.}}}\label{sys1fig2}
\vspace{-1.5em}
\end{figure}
The TDMA scheme is adopted for different groups as shown in Fig.~\ref{sys1fig2}, i.e.,
users in each group will be assigned with a fraction of time to use the whole bandwidth.
Let $x_i$ denote the fraction of time allocated to users in group $i$.
In Fig.~\ref{sys1fig2}, $T_s$ is the duration of each time slot.
In each time slot, users in group $i$ transmit with time duration $x_iT_s$.
To ensure time sharing among $N$ groups, we have
\begin{equation}
\sum_{i=1}^N x_i =1.
\end{equation}
Thus, according to \cite{8006982}, the data rate  of user $j$  in group $i$ can be expressed as
\begin{equation} \label{sys1eq2_2}
r_{ij}=x_i \bar r_{ij},  \quad \forall i \in \mathcal N, j\in\mathcal J_i,
\end{equation}
where
\begin{equation}\label{sys1eq2}
\bar r_{ij}=B  \log_2 \left( 1+\frac {p_{ij} h_{ij} } {\sigma^2B +\sum_{l=j+1}^{M_i}p_{il} h_{il} } \right), \quad \forall i \in \mathcal N, j\in\mathcal J_i
\end{equation}
is the Shannon channel capacity of user $j$ in group $i$.

Note that the BS detects the messages of $M_i$ users via NOMA technique, i.e., to detect the message of user $j$, the BS first detects the message of strong user $l\leq j$ and then detects the message of weak user $l>j$ with SIC \cite{7022998,8294259,8320533}.
As a result the Shannon channel capacity of user $j$   in group $i$ can be presented as \eqref{sys1eq2}.
%
%
\subsection{Offloading Model}
The data transmission time for users in group $i$ is denoted by $t_i$.
To meet the uploaded data demand, we have
\begin{equation}\label{sys1eq2_1}
d_{ij} =  {t_i}{r_{ij}}, \quad \forall i \in \mathcal N, j \in\mathcal J_i,
\end{equation}
where $r_{ij}$ is data rate of user $j$ in group $i$ defined in \eqref{sys1eq2_2}.
According to Fig.~\ref{sys1fig2}, the time for user $j$ in group $i$ to transmit with power $p_{ij}$ is $x_it_i$.
To offload $d_{ij}$ bits in time duration $t_i$ with time sharing fraction $x_i$, the energy consumption at user $j$ in  group $i$ is
\begin{equation}\label{sys1eq6}
E_{ij}^{\text{Off}} = p_{ij}x_it_{i}, \quad \forall i \in \mathcal N, j \in \mathcal J_i.
\end{equation}


In the offloading process, the execution time contains the data transmission time and processing time at the cloud.
Let $F$ in CPU cycles per second denote the computation capacity of the edge cloud.
The total computation capacity is split among users, and  the computation capacity at the edge cloud  allocated for the offloaded job of user $j$ in group $i$ is denoted by $f_{ij}$.
Due to limited computation capacity at the edge, we can obtain
\begin{equation}\label{sys1eq6_1}
\sum_{i=1}^N \sum_{j=1}^{M_i}f_{ij} \leq F.
\end{equation}
With transmission time $t_{i}$, the offloading time for user $j$ in group $i$
is
\begin{equation}\label{sys1eq6_2}
T_{ij}^{\text{Off}} =  {t_{i}} +\frac{C_{ij}d_{ij}}{f_{ij}}, \quad \forall i \in \mathcal N, \forall j \in \mathcal J_i.
\end{equation}

\vspace{-1em}
\subsection{Problem Formulation}
\vspace{-0.5em}
Denote $T$ as the completion time of all users.
Now, we are ready to formulate the energy efficient resource allocation problem for the NOMA-enabled MEC network as:
\begin{subequations}\label{sys1min1}
\begin{align}
\mathop {\min}_{\pmb d, \pmb x, \pmb t, \pmb f, \pmb p,T}\;
 \quad& \:  {\color{myc1}{\omega T + (1-\omega)\left(\sum_{i=1}^N \sum_{j=1}^{M_i} ( p_{ij}x_i t_i +C_{ij}Q_{ij}(R_{ij}-d_{ij})) \right)}}
  \\
\textrm{s.t.}\qquad&  \frac{C_{ij}(R_{ij}-d_{ij})  }{F_{ij}} \leq T, \quad \forall i\in\mathcal N, j\in\mathcal J_i\\
&  {t_{i}} +\frac{C_{ij}d_{ij}}{ f_{ij}} \leq T, \quad \forall i\in\mathcal N, j\in\mathcal J_i \\
&  Bx_it_i\log_2 \left( 1+\frac {p_{ij} h_{ij} } {\sigma^2B +\sum_{l=j+1}^{M_i}p_{il} h_{il} } \right)=  d_{ij},\! \quad \forall i\in\mathcal N, j\in\mathcal J_i \\
&\sum_{i=1}^N x_i = 1 \\
& \sum_{i=1}^N \sum_{j=1}^{M_i} f_{ij} \leq F  \\
&0\leq   p_{ij}\leq P_{ij},  \quad \forall i\in\mathcal N, j\in\mathcal J_i\\
&0\leq d_{ij}\leq R_{ij}, \quad \forall i\in\mathcal N, j\in\mathcal J_i
\label{sys1min1i}\\
&  x_i, t_i,  f_{ij} \geq 0, \quad \forall i\in\mathcal N, j\in\mathcal J_i,
\end{align}
\end{subequations}
where $\pmb d=[d_{11},\cdots,d_{1M_1},\cdots, d_{NM_N}]$, $\pmb x=[x_1, \cdots,x_N]$,  $\pmb t=[t_1, \cdots,t_N]$, $\pmb f=[f_{11},\cdots,$
$f_{1M_1},\cdots,f_{NM_N}]$, $\pmb p=[p_{11},\cdots,p_{1M_1},\cdots,p_{NM_N}]$,
$\omega\in[0,1]$ is a constant parameter
and $P_{ij}$ is the maximal transmission power of user $j$ in group $i$.
In the objective function (\ref{sys1min1}a), $\omega$ is used
to characterize the tradeoff of the completion time $T$ and total energy consumption, $\sum_{i=1}^N \sum_{j=1}^{M_i} ( p_{ij}x_i t_i +C_{ij}Q_{ij}(R_{ij}-d_{ij}))$, including both offloading energy and local computing energy.
Constraints (\ref{sys1min1}b) reflect that the execution time of the local tasks for all users should not exceed the prescribed completion time,
while constraints (\ref{sys1min1}c) mean that the execution time of the offloaded tasks (including the transmission time)  for all users should not exceed the completion time.
The offloaded data demand should be satisfied as stated in constraints (\ref{sys1min1}d).
The time division constraint is shown in (\ref{sys1min1}e), and the edge cloud capacity sharing constraint is given in (\ref{sys1min1}f).
Constraints (\ref{sys1min1}g) and constraints \eqref{sys1min1i} respectively represent the maximal power and offloading data limits of all users.
Due to the nonconvex objective function (\ref{sys1min1}a) and constraints (\ref{sys1min1}c)-(\ref{sys1min1}d), Problem (\ref{sys1min1}) is nonconvex, which is hard to obtain the globally optimal solution.

\section{Energy Efficient Resource Allocation}

\begin{figure}[t]
\centering
\includegraphics[width=5in]{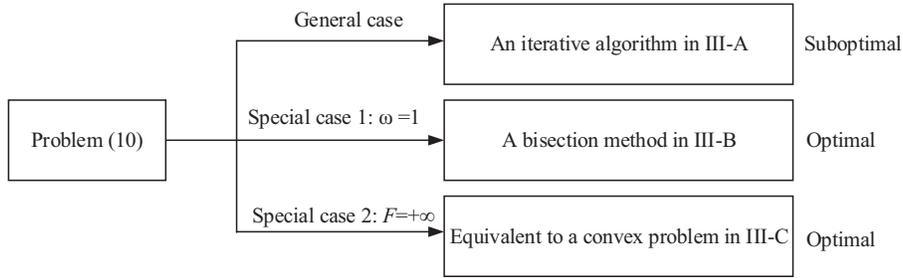}
\vspace{-1.5em}
\caption{{\color{myc1}{Proposed approach for solving Problem (\ref{sys1min1}).}}} \label{strcuter}
\vspace{-1.5em}
\end{figure}

{\color{myc1}{
In this section, we first transform Problem (\ref{sys1min1}) into an equivalent problem, which can be solved via an iterative algorithm with low complexity.
Next, we also provide an effective algorithm to obtain the optimal solution of Problem (\ref{sys1min1}) with $\omega=1$.
We then show that the optimal solution of Problem (\ref{sys1min1}) can be obtained by solving an equivalent convex problem for the case with infinite cloud capacity.
Finally, the algorithm analysis is provided.
The proposed process for solving Problem (\ref{sys1min1}) is summarized in Fig. \ref{strcuter}.}}

\subsection{Iterative Algorithm}
To simplify Problem (\ref{sys1min1}), we provide the following lemma.
\begin{lemma}
Problem (\ref{sys1min1}) is equivalent  to the following problem:
\begin{subequations}\label{ener3min1}
\begin{align}
\mathop{\min}_{\pmb d, \pmb x, \pmb \tau, \pmb f, \pmb p, T}\;
 \quad&   \omega T + (1-\omega)\left(\sum_{i=1}^N \sum_{j=1}^{M_i} ( p_{ij}\tau_i +C_{ij}Q_{ij}(R_{ij}-d_{ij})) \right)
  \\
\textrm{s.t.}\quad\qquad \!\!\!\!\!
&  \frac{C_{ij}(R_{ij}-d_{ij})  }{F_{ij}} \leq T, \quad \forall i\in\mathcal N, j\in\mathcal J_i\\
& \frac{\tau_i}{x_i} +\frac{C_{ij}d_{ij}}{f_{ij}}\leq T, \quad \forall i\in\mathcal N, j\in\mathcal J_i\\
& B\tau_i\log_2 \left( 1+\frac { \sum_{l=j}^{M_i} p_{il} h_{il} } {\sigma^2B   } \right)\geq \sum_{l=j}^{M_i} d_{il}  ,  \quad \forall i\in\mathcal N, j \in\mathcal J_i \\
&\sum_{i=1}^N x_i = 1 \\
& \sum_{i=1}^N \sum_{j=1}^{M_i} f_{ij} \leq F  \\
&0\leq   p_{ij}\leq P_{ij},  \quad \forall i\in\mathcal N, j\in\mathcal J_i\\
&0\leq d_{ij}\leq R_{ij}, \quad \forall i\in\mathcal N, j\in\mathcal J_i\\
&  x_i, \tau_i,  f_{ij} \geq 0, \quad \forall i\in\mathcal N, j\in\mathcal J_i,
\end{align}
\end{subequations}
where $\pmb \tau=[\tau_1, \cdots, \tau_N]$.
\end{lemma}

\itshape \textbf{Proof:}  \upshape
Refer to Appendix A.
\hfill $\Box$

Compared with Problem (\ref{sys1min1}), the equivalent Problem (\ref{ener3min1}) is simplified since constraints (\ref{ener3min1}d) are convex with respect to (w.r.t.) power $\pmb p$ now.
Due to the nonconvex objective function (\ref{ener3min1}a) and constraints (\ref{ener3min1}c) and (\ref{ener3min1}d), it is generally hard to obtain the globally optimal solution of nonconvex Problem (\ref{ener3min1}).
To obtain a suboptimal solution of Problem (\ref{ener3min1}), we present an iterative algorithm.


Before solving Problem \eqref{ener3min1}, several properties are provided as follows.

\begin{lemma}\label{lem2}
With fixed time allocation, computation capacity allocation and completion time  $(\pmb \tau, \pmb f, T)$,  Problem (\ref{ener3min1}) is convex w.r.t. $(\pmb d, \pmb x, \pmb p)$.
With given offloaded data, time sharing factor and power control $(\pmb d, \pmb x, \pmb p)$, Problem (\ref{ener3min1}) is convex w.r.t. $(\pmb \tau, \pmb f, T)$.
\end{lemma}

Since Lemma \ref{lem2} can be easily proved according to the fact that $\log (x)$ is concave and $\frac{1}{x}$ is convex, the proof is omitted.
Lemma \ref{lem2} shows that problem \eqref{ener3min1} is a convex problem with fixed $(\pmb \tau, \pmb f, T)$ or $(\pmb d, \pmb x, \pmb p)$, which can be solved by using the iterative algorithm.


\begin{lemma}\label{lem3}
Constraints (\ref{ener3min1}c) hold with equality at the optimal solution for Problem (\ref{ener3min1}):
\begin{equation}
\frac{\tau_i^*}{x_i^*} +\frac{C_{ij}d_{ij}^*}{f_{ij}^*} = T^*, \quad \forall i\in\mathcal N, j\in\mathcal J_i,
\end{equation}
where $d_{ij}^*,x_{i}^*, \tau_i^*,f_{ij}^*$ and $T^*$ denote the optimal solution.
\end{lemma}

\itshape \textbf{Proof:}  \upshape
Refer to Appendix B.
\hfill $\Box$

Lemma \ref{lem3} shows that utilizing the total completion time is optimal.
This is because the edge computation capacity is limited and long execution time at the edge is always computation capacity saving.

To solve nonconvex Problem \eqref{ener3min1}, we propose an iterative algorithm via optimizing $(\pmb \tau, \pmb f, T)$ with fixed $(\pmb d, \pmb x, \pmb p)$ and solving $(\pmb d, \pmb x, \pmb p)$ with  fixed $(\pmb \tau, \pmb f, T)$.

With fixed offloaded data, time sharing factor and power control $(\pmb d, \pmb x, \pmb p)$, Problem \eqref{ener3min1} becomes the following convex problem:
\begin{subequations}\label{ener3min2}
\begin{align}
\mathop{\min}_{ \pmb \tau, \pmb f, T}\;
 \quad&   \omega T + (1-\omega) \sum_{i=1}^N \sum_{j=1}^{M_i}   p_{ij}\tau_i
  \\
\textrm{s.t.}\quad\qquad \!\!\!\!\!\!\!\!\!
& \frac{\tau_i}{x_i} +\frac{C_{ij}d_{ij}}{f_{ij}}\leq T, \quad \forall i\in\mathcal N, j\in\mathcal J_i\\
& \sum_{i=1}^N \sum_{j=1}^ {M_i}f_{ij} \leq F  \\
& \tau_i \geq T_i,  f_{ij} \geq 0, T \geq \bar T, \quad \forall i\in\mathcal N, j\in\mathcal J_i,
\end{align}
\end{subequations}
where
\begin{equation}
T_i=\max_{j\in\mathcal J_i} \frac {\sum_{l=j}^{M_i} d_{il}}{B\log_2 \left( 1+\frac { \sum_{l=j}^{M_i} p_{il} h_{il} } {\sigma^2B   } \right)},
\bar T=\max_{i\in\mathcal N, j\in\mathcal J_i} \frac{C_{ij}(R_{ij}-d_{ij})  }{F_{ij}},
\quad \forall i\in\mathcal N.
\end{equation}

\begin{lemma}
The optimal solution of Problem \eqref{ener3min2} is
\begin{equation}\label{ener3min2eq1_1}
\tau_i^*=T_i,  f_{ij}^*=\frac{C_{ij}d_{ij}x_i}{  T^* x_i-T_i},
T^* =\max \{\bar T, \tilde T\},\quad \forall i\in\mathcal N, j\in\mathcal J,
\end{equation}
where  $\tilde T$ is the solution to
\begin{equation}\label{ener3min2eq1_2}
\sum_{i=1}^N \sum_{j=1}^ {M_i}\frac{C_{ij}d_{ij}x_i}{\tilde T x_i-T_i}=F.
\end{equation}
\end{lemma}

\itshape \textbf{Proof:}  \upshape
Refer to Appendix C.
\hfill $\Box$

With fixed time allocation, computation capacity allocation and completion time  $(\pmb \tau, \pmb f, T)$, the total energy minimization Problem \eqref{ener3min1} is simplified as:
\begin{subequations}\label{ener3min5}
\begin{align}
\mathop{\min}_{ \pmb d, \pmb x, \pmb p}\;
 \quad&  \sum_{i=1}^N \sum_{j=1}^{M_i} ( p_{ij}\tau_i +C_{ij}Q_{ij}(R_{ij}-d_{ij}))
  \\
\textrm{s.t.}\quad\qquad \!\!\!\!\!\!\!\!\!
& \frac{\tau_i}{x_i} +\frac{C_{ij}d_{ij}}{f_{ij}}\leq T, \quad \forall i\in\mathcal N, j\in\mathcal J_i\\
&  B\tau_i\log_2 \left( 1+\frac { \sum_{l=j}^{M_i} p_{il} h_{il} } {\sigma^2B   } \right)\geq \sum_{l=j}^{M_i} d_{il}  ,  \quad \forall i\in\mathcal N, j \in\mathcal J_i \\
&\sum_{i=1}^N x_i = 1 \\
&0\leq   p_{ij}\leq P_{ij},  \quad \forall i\in\mathcal N, j\in\mathcal J_i\\
& D_{ij} \leq d_{ij} \leq R_{ij}, x_i\geq 0, \quad \forall i\in\mathcal N, j\in\mathcal J_i,
\end{align}
\end{subequations}
where $D_{ij}=\max\left\{\frac{C_{ij}R_{ij}-TF_{ij}}{C_{ij}},0\right\}$.
Due to the convexity, Problem \eqref{ener3min5} can be effectively solved by the dual method, i.e., iteratively optimizing primal variables with fixed Lagrange multipliers and solving Lagrange multipliers with optimized primal variables.
The details of solving Problem \eqref{ener3min5} with the dual method can be found in Appendix D.


\begin{algorithm}[h]
\caption{: Iterative Algorithm}
\begin{algorithmic}[1]
\STATE
 Initialize a feasible solution $(\pmb d^{(0)}, \pmb x^{(0)}, \pmb \tau^{(0)}, \pmb f^{(0)}, \pmb p^{(0)}, T^{(0)})$ of Problem (\ref{sys1min1})  and set $l=0$.
 \REPEAT
\STATE With given offloading data, time sharing factor and power control $(\pmb d^{(l)}, \pmb x^{(l)}, \pmb p^{(l)})$, obtain the optimal $(\pmb \tau^{(l+1)}, \pmb f^{(l+1)}, T^{(l+1)})$ of Problem \eqref{ener3min2}.
\STATE With given time allocation, computation capacity allocation and completion time $(\pmb \tau^{(l+1)}, \pmb f^{(l+1)}, T^{(l+1)})$, obtain the  optimal $(\pmb d^{(l+1)}, \pmb x^{(l+1)}, \pmb p^{(l+1)})$  of Problem \eqref{ener3min5}.
\STATE Set $l=l+1$.
\UNTIL {Convergence}
\STATE Output $\pmb d^*=\pmb d^{(l)}$, $\pmb x^*=\pmb x^{(l)}$,  $\pmb f^*=\pmb f^{(l)}$, $\pmb p^*=\pmb p^{(l)}$, $T^*=T^{(l)}$, $t_{i}^*=\frac{\tau_i^{(l)}}{x_i^{(l)}}$, $\forall i \in \mathcal N$.
\end{algorithmic}
\end{algorithm}

By iteratively solving Problem \eqref{ener3min2} and  Problem \eqref{ener3min5}, the algorithm that solves Problem (\ref{sys1min1}) is given in Algorithm~1.
Since the optimal solution of  Problem \eqref{ener3min2} or  \eqref{ener3min5} is obtained in each step, the objective value of   Problem (\ref{sys1min1}) is nonincreasing in each step.
Moreover, the objective value of Problem (\ref{sys1min1})  is lower bounded by zero.
Thus, Algorithm~1 must converge.

\subsection{Completion Time Minimization with $\omega=1$}
{\color{myc1}{In this section, we consider the completion time minimization Problem (\ref{sys1min1}) with $\omega=1$, i.e., the objective function is  $T$.
To solve Problem (\ref{sys1min1}) with $\omega=1$, we have the following lemma.

\begin{lemma}
For the optimal solution $(\pmb d^*, \pmb x^*, \pmb t^*, \pmb f^*, \pmb p^*, T^*)$ of Problem (\ref{sys1min1}) with $\omega=1$,  Problem (\ref{sys1min1}) with $\omega=1$ and $T<T^*$ does not have a feasible solution (i.e., it is infeasible), and  Problem (\ref{sys1min1}) with $\omega=1$ and $T>T^*$ always has a feasible solution (i.e., it is feasible).
\end{lemma}

\itshape \textbf{Proof:}  \upshape
Assume that Problem (\ref{sys1min1}) with $\omega=1$ and $\bar T<T^*$ is feasible, and the feasible solution is $(\bar {\pmb d}, \bar{\pmb x}, \bar{\pmb t}, \bar{\pmb f})$.
Then, the solution $(\bar {\pmb d}, \bar{\pmb x}, \bar{\pmb t}, \bar{\pmb f}, \bar T)$ is feasible with lower value of the objective function than solution $(\pmb d^*, \pmb x^*, \pmb t^*, \pmb f^*, \pmb p^*, T^*)$, which contradicts the fact that $(\pmb d^*, \pmb x^*, \pmb t^*, \pmb f^*, \pmb p^*, T^*)$ is the optimal solution.

For Problem (\ref{sys1min1}) with $\omega=1$ and $\bar T>T^*$, we can always construct a feasible solution $(\pmb d^*, \pmb x^*, \pmb t^*, \pmb f^*, \pmb p^*, \bar T )$ to Problem (\ref{sys1min1}) with $\omega=1$ by checking constraints (\ref{sys1min1}b)-(\ref{sys1min1}i). \hfill $\Box$

According to Lemma 5, we can utilize the bisection method to solve Problem (\ref{sys1min1}) with $\omega=1$.
Denote $T_{\min}=0$, $T_{\max}=\max_{i\in\mathcal N, j \in\mathcal J_i} \frac{C_{ij} R_{ij}}{F_{ij}}$.
If  $T>T_{\max}$, Problem (\ref{sys1min1}) with $\omega=1$ is always feasible by setting $d_{ij}=x_i=t_i=f_{ij}=p_{ij}=0$.
As a result, the optimal $T^*$ of Problem (\ref{sys1min1}) with $\omega=1$ must lie in the interval $(T_{\min}, T_{\max})$.
At each step the bisection method divides the interval in two by computing the midpoint $T_{\text{mid}}=(T_{\min}+T_{\max})/2$. There are now only two possibilities: 1) if Problem (\ref{sys1min1}) with $\omega=1$ and $T=T_{\text{mid}}$ is feasible,  we have $T^*\in(T_{\min},T_{\text{mid}}]$,
2) if Problem (\ref{sys1min1}) with $\omega=1$ and $T=T_{\text{mid}}$ is infeasible, we have   $T^*\in(T_{\text{mid}},T_{\max})$.
The bisection method selects the subinterval that is guaranteed to be a bracket as the new interval to be used in the next step. In this way an interval that contains the optimal $T^*$ is reduced in width by 50\% at each step. The process is continued until the interval is sufficiently small.
}}

For each given $T$, we solve a feasibility problem with constraints (\ref{sys1min1}b)-(\ref{sys1min1}i).
According to constraints (\ref{sys1min1}b), we have
\begin{equation}\label{time2eq1}
d_{ij}\geq D_{ij}=\max\left\{\frac{C_{ij}R_{ij}-TF_{ij}}{C_{ij}},0\right\},\quad \forall i\in\mathcal N, j\in\mathcal J_i.
\end{equation}
Based on constraints (\ref{sys1min1}c)-(\ref{sys1min1}d) about $d_{ij}$, we find that it is optimal to offload the smallest data to minimize the completion time, i.e., $d_{ij}=D_{ij}$.
Substituting $d_{ij}=D_{ij}$ into Problem (\ref{sys1min1}) with objective function $T$, {\color{myc1}{the feasibility set becomes}}
\vspace{-0.5em}
\begin{subequations}\label{time2min1}\vspace{-0.5em}
\begin{align}
%
 \text{find} \quad\: & \pmb x, \pmb t, \pmb f, \pmb p
  \\
\textrm{s.t.}\quad\qquad \!\!\!\!\!\!\!\!\!
& t_{i}+\frac{C_{ij}D_{ij}}{ f_{ij}} \leq T, \quad \forall i\in\mathcal N, j\in\mathcal J_i\\
& Bx_it_i\log_2 \left( 1+\frac {p_{ij} h_{ij} } {\sigma^2B +\sum_{l=j+1}^{M_i}p_{il} h_{il} } \right)=  D_{ij},\! \quad \forall i\in\mathcal N, j\in\mathcal J_i\\
& \sum_{i=1}^N x_i=1  \\
& \sum_{i=1}^N \sum_{j=1}^{M_i} f_{ij} \leq F  \\
&  0\leq p_{ij}\leq P_{ij}, \quad \forall i\in\mathcal N, j\in\mathcal J_i\\
&   x_i, t_i, f_{ij}\geq 0, \quad \forall i\in\mathcal N, j\in\mathcal J_i.
\end{align}
\end{subequations}

Since constraints (\ref{time2min1}c) are nonconvex, {\color{myc1}{set (\ref{time2min1}) is nonconvex}}.
To concur the nonconvexity of (\ref{time2min1}), we treat time vector $\pmb t$ and power vector $\pmb p$ as intermediate variables and obtain the following theorem.
\begin{theorem}
Feasibility set (\ref{time2min1})  is equivalent  to the following convex set:
\vspace{-0.5em}
\begin{subequations}\label{AppBmin2}\vspace{-0.5em}
\begin{align}
 \text{find} \quad\: & \pmb x, \pmb f
  \\
\textrm{s.t.}\quad\qquad \!\!\!\!\!\!\!\!\!
& \frac{\bar T_{i}}{x_i} +\frac{C_{ij}D_{ij}}{ f_{ij}} \leq T, \quad \forall i\in\mathcal N, j\in\mathcal J_i \\
&\sum_{i=1}^N x_i = 1 \\
& \sum_{i=1}^N \sum_{j=1}^2 f_{ij} \leq F  \\
& x_i, f_{ij} \geq 0, \quad \forall i\in\mathcal N, j\in\mathcal J_i,
\end{align}
\end{subequations}
where
\vspace{-0.5em}
\begin{equation}\label{AppThe2eq9}\vspace{-0.5em}
\bar T_i =\max_{j\in\mathcal J_i} T_{ij}, \quad \forall i \in \mathcal N,
\end{equation}
and {\color{myc1}{$T_{ij}$ is the solution to the following equation
\vspace{-0.5em}
\begin{equation}\label{AppThe2eq8}\vspace{-0.5em}
P_{ij}=\frac 1 {h_{ij}}{\left(2^{\frac{D_{ij}}{B T_{ij}}}-1\right)} \sum_{l=j+1}^{M_i} 2^{\frac{\sum_{k=j+1}^{l-1} D_{ik}}{BT_{ij}}}\left({2^{\frac{D_{il}}{BT_{ij}}}} -1\right)\sigma^2B +\frac 1 {h_{ij}}\left({2^{\frac{D_{ij}}{BT_{ij}}}} -1\right)\sigma^2B.
\end{equation}}}
\end{theorem}

\itshape \textbf{Proof:}  \upshape
Refer to Appendix E.
\hfill $\blacksquare$

\begin{lemma}
The necessary and sufficient conditions for that set \eqref{AppBmin2} is non-empty are
\begin{equation}\label{time2eq3}
\sum_{i=1}^N \bar T_i \leq T,
\end{equation}
and
\begin{equation}\label{time2eq3_2}
\frac{\left( { \sum_{i=1}^N \sqrt{  {\bar T_i\sum_{j=1}^{M_i}C_{ij}D_{ij} } }} \right)^2}
{T\left(T-\sum_{i=1}^N \bar T_i\right)}
+\sum_{i=1}^N \sum_{j=1}^{M_i} \frac{C_{ij} D_{ij}}{T} \leq F.
\end{equation}
\end{lemma}

\itshape \textbf{Proof:}  \upshape
Refer to Appendix F.
\hfill $\Box$

\begin{algorithm}[h]
\caption{: Minimal Completion Time}
\begin{algorithmic}[1]
 \STATE Initialize $T_{\min}=0$, $T_{\max}=\max_{i\in\mathcal N, j \in\mathcal J_i} \frac{C_{ij} R_{ij}}{F_{ij}}$, and the tolerance $\epsilon$.
 \STATE {\color{myc1}{Set $T=\frac{T_{\min}+T_{\max}}{2}$}}, and calculate $D_{ij}$ and $\bar T_i$ according to (\ref{time2eq1}) and (\ref{AppThe2eq9}), respectively.
 \STATE Check the feasibility conditions (\ref{time2eq3}) and (\ref{time2eq3_2}). If (\ref{AppBmin2}) has  a feasible solution, {\color{myc1}{set $T_{\max}=T$}}. Otherwise, {\color{myc1}{set $T=T_{\min}$}}.
 \STATE If $(T_{\max}-T_{\min})/T_{\max}\leq \epsilon$, terminate. Otherwise, go to step 2.
\end{algorithmic}
\end{algorithm}

Based on Lemma 6, the algorithm for obtaining the minimal completion time is summarized in Algorithm 2.

\subsection{Energy Efficient Optimization with Infinite Cloud Capacity}
In this section, we consider the case that the computation capacity at the BS is infinite, i.e., $F$ is so large that the computation time at the BS is neglected.
To solve Problem (\ref{ener3min1}), we can show that it can be transformed into an equivalent convex one, which is stated by the following theorem.
\begin{theorem}
For infinite cloud capacity,   Problem (\ref{ener3min1})  is equivalent  to the following convex problem:
\vspace{-0.5em}
\begin{subequations}\label{ener3min1_2}\vspace{-0.5em}
\begin{align}
\mathop{\min}_{\pmb d, \pmb \tau,  \pmb q, T}\;
 \quad&  \omega T + (1-\omega)\left(\sum_{i=1}^N \sum_{j=1}^{M_i} ( q_{ij}  +C_{ij}Q_{ij}(R_{ij}-d_{ij})) \right)
  \\
\textrm{s.t.}\quad\qquad \!\!\!\!\!\!\!\!\!
& B\tau_i\log_2 \left( 1+\frac { \sum_{l=j}^{M_i} q_{il} h_{il} } {\sigma^2B \tau_i  } \right)\geq \sum_{l=j}^{M_i} d_{il}  ,  \quad \forall i\in\mathcal N, j \in\mathcal J_i \\
&\sum_{i=1}^N \tau_i \leq T \\
&   q_{ij}\leq P_{ij}\tau_i, \quad \forall i\in\mathcal N, j\in\mathcal J_i\\
& D_{ij} \leq d_{ij}\leq R_{ij}, \quad \forall i\in\mathcal N, j\in\mathcal J_i\\
&   \tau_i, q_{ij} \geq 0, \quad \forall i\in\mathcal N, j\in\mathcal J_i,
\end{align}
\end{subequations}
where $\pmb q=[q_{11},\cdots,q_{1M_1},\cdots,q_{NM_N}]$, and we set $  B\tau_i\log_2 \left( 1+\frac { \sum_{l=j}^{M_i} q_{il} h_{il} } {\sigma^2B \tau_i  } \right)=0$ for the case $\tau_i=0$. 
\end{theorem}

\itshape \textbf{Proof:}  \upshape
Refer to Appendix G.
\hfill $\blacksquare$

Theorem 2 indicates that the energy efficient resource allocation problem with infinite cloud capacity is equivalent to a convex problem, of which the globally optimal solution can be effectively obtained \cite{boyd2004convex}.

\begin{lemma}\label{maxtime}
For Problem \eqref{ener3min1_2}, it is optimal to transmit with maximal time, i.e.,  $\sum_{i=1}^N \tau_i^*=T^*$.
\end{lemma}

\itshape \textbf{Proof:}  \upshape
Refer to Appendix H.
\hfill $\Box$

According to Lemma \ref{maxtime}, transmitting with maximal time is always energy efficient.
The reason is that, as the transmission time increases,
the required power decreases and then the product of time and
power, which can be viewed as the consumed energy, also
decreases.

{\color{myc1}{
\subsection{Algorithm Analysis}
To solve the general energy efficient resource allocation Problem \eqref{sys1min1} by using Algorithm 1, the major complexity in each step lies in solving the offloaded data, time sharing factor and power control of Problem (\ref{ener3min5}).
From Appendix D, the complexity  of solving Problem (\ref{ener3min5}) with the dual method is $\mathcal O(L_{\text{dm1}}M)$ \cite{6891348}, where $L_{\text{dm1}}$ is the number of iterations of the dual method.
As a result, the total complexity of the proposed Algorithm 1 is $\mathcal O(L_{\text{it}}L_{\text{dm1}}N)$, where $L_{\text{it}}$ is the number of iterations for iteratively optimizing $(\pmb \tau, \pmb f, T)$ and $(\pmb d, \pmb x, \pmb p)$.

For the spacial case with $\omega=1$, the optimal solution of Problem \eqref{sys1min1} is obtained by using Algorithm 2.
According to Algorithm 2, the main complexity in each step lies in checking the feasibility conditions (\ref{time2eq3}) and (\ref{time2eq3_2}), which involves complexity $\mathcal O(M)$.
As a result, the total complexity of Algorithm 2 is $\mathcal O(M\log2(1/\epsilon))$, where $\mathcal O(\log2(1/\epsilon))$ is the complexity of the bisection method with accuracy $\epsilon$.

For the special case with infinite cloud complexity, Problem \eqref{sys1min1} is equivalent to a convex Problem (\ref{ener3min1_2}) according to Theorem~2, which can be effectively solved via the dual method as in Appendix D.
Due to the fact that the dimension of the variables in Problem (\ref{ener3min1_2}) is $\mathcal O (M)$, the complexity  of solving Problem (\ref{ener3min1_2}) is $\mathcal O(L_{\text{dm2}}M)$ \cite{6891348}, where $L_{\text{dm2}}$ is the number of iterations of the dual method.


As a result, it is observed that the complexity of the proposed algorithms  grows linearly with the number of all users $M$.
Besides, the proposed algorithms are all centralized ones and the BS needs to collect the information from all users.
To implement the proposed algorithms, all the users first transmit the pilot sequence to the BS, and the BS obtains the uplink channel conditions of all users, which involves overhead $M$.
Then, all users need to upload the information about the local computation capabilities, power limits and input data sizes to the BS, which involves overhead $3M$.
Thus, the total overhead at the BS is $4M$, which grows linearly with the number of all users.

If users enter or exit the system, the BS can re-group users according to the channel gains and run the proposed algorithms to obtain the offloaded data, the time sharing factor, time allocation, computation capacity allocation, and transmission power.
Then, the BS transmits the information about the offloaded data $\pmb d$,  time sharing factor $\pmb x$, time allocation $\pmb t$,  transmission power $\pmb p$ and re-group information to all $M$ users, which involves overhead $5M$.
}}

\vspace{-0.1em}
\section{Numerical Results}
\vspace{-0.5em}
In this section, numerical results are presented  to evaluate the performance of the proposed algorithm.
The NOMA-enabled MEC network consists of $M=30$ users.
The path loss model is $128.1+37.6\log_{10} d$ ($d$ is in km),
 the standard deviation of shadow fading is $4$ dB
 and the small-scale channel gain is exponentially distributed with parameter 1 \cite{access2010further}.
In addition, the bandwidth of the network is $B=10$ MHz, and the noise power density is  $\sigma^2=-169$ dBm/Hz.
For MEC parameters, the required number of CPU cycles per bit is  set to follow equal distribution $C_{ij}\in[500, 1500]$ cycles/bit.
The CPU computation of each user is set as the same $F_{ij}=1$ GHz and the local computation energy per cycle for each user is also set as equal $Q_{ij}=10^{-10}$ J/cycle for all $i \in \mathcal N$,  $j\in\mathcal J_i$.
We consider equal data size and equal maximal transmission power for all users, i.e., $d_{ij}=D$ and $P_{ij}=P$, for all $i \in \mathcal N$, $j \in \mathcal J_i$.
Unless specified otherwise, the system parameters are set as  maximal transmission power $P=1$ dBm, offloaded data $D= 100$ Kbits, and edge cloud capacity $F=2\times 10^{10}$ cycles/s.

We compare  the proposed NOMA scheme with the TDMA  and FDMA schemes in \cite{8006982}, where all users are respectively allocated with different time slots and frequency bands,
 the exhaustive search method to obtain a near globally optimal solution of Problem (\ref{sys1min1}) (labelled as `NOMA-EXH'), which refers to the proposed Algorithm 1 with 1000 initial starting points,
and  {\color{myc2}{the exhaustive search method with equal time sharing factor (labelled as `NOMA-ET'), which refers to the proposed Algorithm 1 with 1000 initial starting points and fixed time sharing factor $x_1=\cdots=x_N=\frac1N$.}}

Due to the decoding complexity and SIC error propagation, we consider the multi-group case, where each group contains two paired users.
For comparison, we also consider the case that there is only one big group with $M$  users (labelled as `NOMA, OG').
To study the influence of user pairing,  we apply three different user-pairing methods in \cite{7273963}.
For strong-strong (SS) pair selection, the user with the strongest channel condition is paired with the one with the second strongest, and
the user with the third strongest is paired with the one with the fourth strongest, and so on.
For strong-weak (SW) pair selection, the user with the strongest channel condition is paired with the user with the weakest, and so on.
For strong-middle (SM) pair selection, the user with the strongest channel condition is paired with the user with the middle strongest user, and so on.

\begin{figure}
\centering
\includegraphics[width=3.4in]{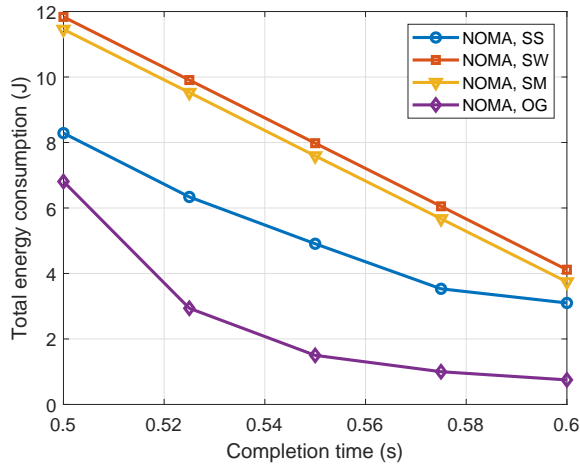}
\vspace{-1em}
\caption{Tradeoff between total energy consumption and completion time with different user pairing methods.}\label{fig17}
\vspace{-1.5em}
\end{figure}

In Fig. \ref{fig17}, we present the tradeoff between total energy consumption and completion time with different user pairing methods.
{\color{myc1}{Note that Fig. \ref{fig17} is obtained by changing the values of parameter $\omega$ in (\ref{sys1min1}a).}}
From Fig. \ref{fig17}, we find that the total energy consumption decreases with  the completion time.
The reason is that transmitting with long time is energy efficient according to Lemma 3.
{\color{myc1}{It is found that NOMA with one big group achieves the best performance.
This is because all the users can form NOMA and the transmission time for all the users in NOMA with one big group is longer than that in NOMA with multiple small groups.
As a result, the energy consumption of NOMA with one big group is lower than that in NOMA with multiple small groups.
However, the decoding complexity is high for NOMA with many users in practical.
It is also observed that SS achieves the best performance among three user pairing methods. This is due to the fact
that users with small difference in channel gains require similar time to offload the same data size, which consequently leads to small completion time and energy consumption.}}
From Fig. \ref{fig17}, it is energy efficient to pair users with similar channel gains.
Due to the performance superiority, the SS pairing method is adopted for NOMA in the following numerical results.

\begin{figure}
\centering
\includegraphics[width=3.4in]{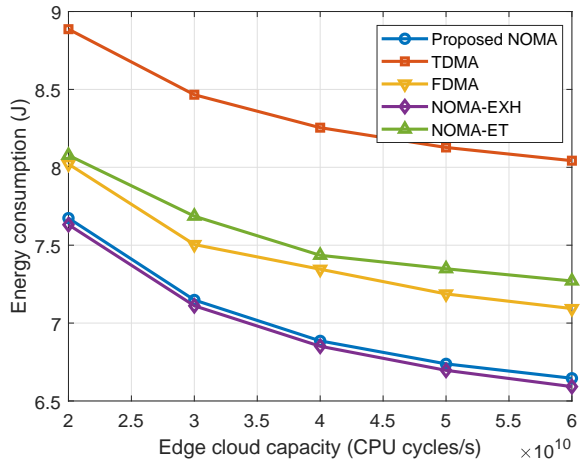}
\vspace{-1em}
\caption{Total energy consumption versus edge cloud capacity with $\omega=0.9$.}\label{fig5}
\vspace{-1.5em}
\end{figure}

The total energy consumption versus edge cloud capacity with $\omega=0.9$ is illustrated in Fig.~\ref{fig5}.
It is observed that the total energy consumption of all schemes decreases with edge cloud capacity since higher edge cloud capacity allows users to offload more data to the BS, resulting lower energy consumption at users.
Besides, the total energy consumption of the proposed NOMA scheme outperforms the conventional TDMA and FDMA schemes, especially when the edge cloud capacity is high.
This is because users in NOMA can simultaneously transmit data by occupying the whole bandwidth, while users in TDMA/FDMA only occupy a fraction of time/frequency resource.
Moreover, the NOMA-EXH algorithm yields the best
performance at the cost of high computation complexity.
The gap between the proposed algorithm for NOMA
and NOMA-EXH is small especially for low edge cloud capacity, which indicates that the proposed algorithm approaches the near globally optimal solution.
{\color{myc2}{According to Fig.~\ref{fig5}, the proposed NOMA outperforms NOMA-ET especially for high edge cloud capacity, which shows the superiority of time allocation.
This is because the data rate between users and the BS  is high  due to time sharing factor optimization in the proposed NOMA, which allows more bits to be uploaded to the BS and decreases the energy consumption at users.}}


\begin{figure}
\centering
\includegraphics[width=3.4in]{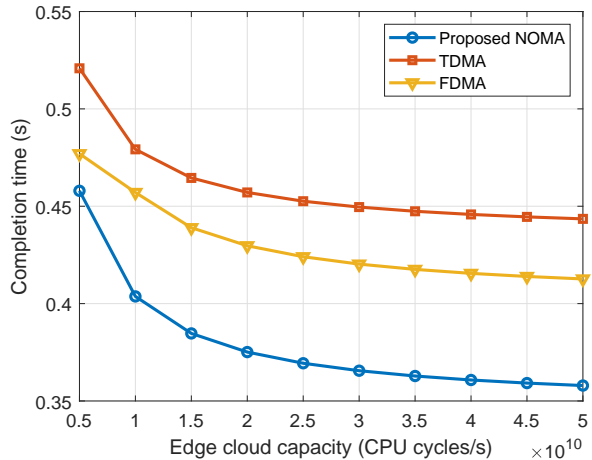}
\vspace{-1em}
\caption{Completion time versus edge cloud capacity with $\omega=1$.}\label{fig1}
\vspace{-1.5em}
\end{figure}

Fig.~\ref{fig1}  depicts the completion time versus edge cloud capacity with $\omega=1$.
{\color{myc1}{It is found that  the completion time decreases with the increase of edge cloud capacity.
This is because high edge cloud capacity leads to low task processing time at the BS.
According to Fig.~\ref{fig1}, FDMA achieves lower completion time than TDMA.
The reason is that the noise power in TDMA is higher than that in FDMA since users utilize frequency division in FDMA.}}
From this figure, we also find that NOMA outperforms TDMA and FDMA in terms of completion time, especially for large edge cloud capacity.
This is because NOMA enables users in each group to simultaneously transmit data to the BS, which is time saving compared to TDMA.
{\color{myc1}{Compared to FDMA, users in NOMA can utilize the whole bandwidth, which is more spectrum efficient and consequently results in lower completion time.}}

\begin{figure}
\centering
\includegraphics[width=3.4in]{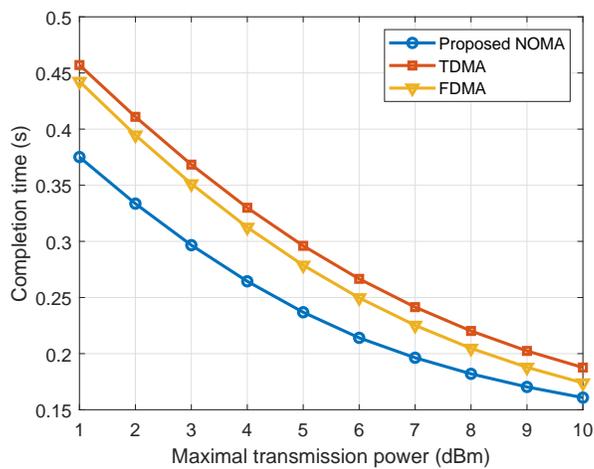}
\vspace{-1em}
\caption{Completion time versus maximal transmission power with $\omega=1$.}\label{fig2}
\vspace{-1.5em}
\end{figure}

In Fig. \ref{fig2},  the completion time versus  maximal transmission power with $\omega=1$ is presented.
It is observed that for all schemes the completion time decreases with the increase of maximal transmission power and the decrease speed is fast for low maximal transmission power region.
The reason is that high maximal transmission power allows users to transmit with high data rate, which reduces the transmission time when offloading data to the BS.
From this figures, it is also found that NOMA outperforms TDMA and FDMA in terms of completion time  especially for low maximal transmission power.
The reason is that NOMA can ensure users in the same group to transmit data to the BS with the same time and frequency resource, which can increase the data rate especially when the transmission power of the user is low.


\begin{figure}
\centering
\includegraphics[width=3.4in]{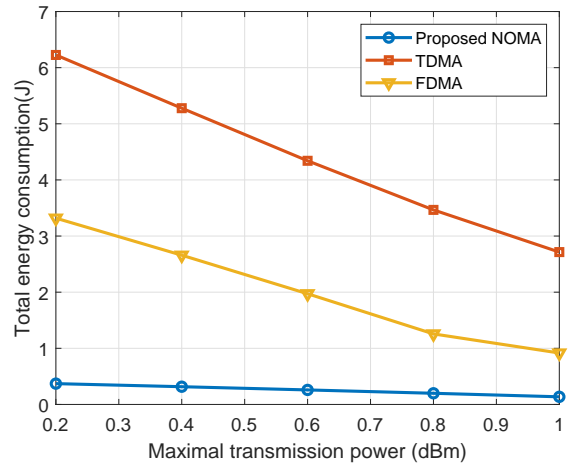}\vspace{-1em}
\caption{Total energy consumption versus maximal transmission power with infinite edge cloud capacity and $\omega=0.9$.}\label{fig15}
\vspace{-1.5em}
\end{figure}

\begin{figure}
\centering
\includegraphics[width=3.4in]{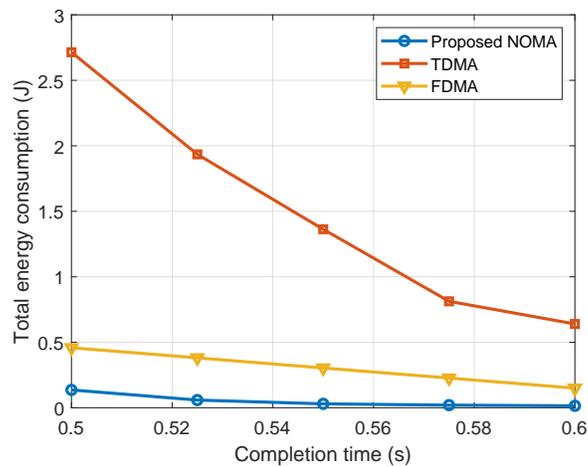}
\vspace{-0.5em}
\caption{Tradeoff between total energy consumption and completion time with infinite edge cloud  capacity.}\label{fig16}
\vspace{-1.5em}
\end{figure}

Fig. \ref{fig15} shows the total energy consumption versus maximal transmission power with infinite edge cloud capacity and $\omega=0.9$.
From this figure, we find that the total energy consumption decreases with the increase of maximal transmission power for all schemes.
This is because high maximal transmission power allows more users to offload data to the BS, which effectively reduces the local computation energy consumption.

The tradeoff between total energy consumption and completion time with infinite edge cloud  capacity is shown in Fig. \ref{fig16}.
It is found that NOMA outperforms TDMA and FDMA in terms of total energy consumption especially for low completion time.
This is because NOMA enables users in each group to simultaneously transmit data to the BS and the transmission time in NOMA is larger than that in TDMA, which results in energy saving compared to TDMA.
{\color{myc1}{For the same completion time, users in NOMA can upload more bits to the BS than FDMA, which reduces the local computation energy.}}
Compared with TDMA and FDMA, NOMA reduces the total energy consumption of all users at the cost of adding computing complexity at the BS due to SIC.

\vspace{-1em}
\section{Conclusion}
\vspace{-0.5em}
{\color{myc1}{In this paper, we have investigated an energy efficient optimization problem to minimize a linear  combination of the completion time and the total energy  for an uplink NOMA-based MEC network.}}
For the general minimization problem, it was first transformed into an equivalent problem, which can be effectively solved by an iterative algorithm with low complexity.
For the special case with only minimizing completion time, we obtained the optimal solution via the bisection method.
For the special case with infinite cloud capacity, we successfully showed that it can be equivalent to a convex problem according to some variable transformations.
Numerical results showed that NOMA outperforms TDMA and FDMA in terms of completion time and total energy consumption, especially for large edge cloud capacity and small maximal transmission power.
Besides, transmitting with long completion time was presented to be energy efficient.
{\color{myc2}{The optimization of user grouping for NOMA-enabled MEC network is left for our future work.}}

\appendices

\vspace{-0.5em}
\section{Proof of Lemma 1}
\setcounter{equation}{0}
\renewcommand{\theequation}{\thesection.\arabic{equation}}
\vspace{-0.5em}

Setting new variable $\tau_i=x_i t_i$ to replace time $t_i$, $\forall i \in \mathcal{N}$, constraints (\ref{sys1min1}c) become (\ref{ener3min1}c).
%
%
Moreover,  constraints (\ref{sys1min1}d)  are equivalent to the following constraints:
\begin{equation}\label{AppCeq1}
B\tau_i\log_2 \left( 1+\frac { \sum_{l=j}^{M_i} p_{il} h_{il} } {\sigma^2B   } \right)= \sum_{l=j}^{M_i} d_{il}  ,  \quad \forall i\in\mathcal N, j \in\mathcal J_i,
\end{equation}
which can be obtained via summing equality constraints (\ref{sys1min1}d).
As a result, Problem (\ref{sys1min1}) is equivalent to Problem (\ref{ener3min1}).
Note that the offloaded data demand constraints (\ref{ener3min1}d) are set with inequality.

The reason is that for the optimal solution to Problem \eqref{ener3min1}, constraints (\ref{ener3min1}d) must hold with equality.
This can be proved by the contradiction method.
Assume that the optimal solution of Problem \eqref{ener3min1} is $(\pmb d^*, \pmb x^*, \pmb \tau^*, \pmb f^*, \pmb p^*, T^*)$ and there exists $i\in\mathcal M$ and $j$ such that (\ref{ener3min1}d) holds with inequality for $(i,j)$.
In this case, we can slightly decrease $p_{ij}^*$ to $p_{ij}'=p_{ij}^*-\epsilon$, where $\epsilon$ is a small positive constant to satisfy constraint  (\ref{ener3min1}d) for $(i,j)$.
If $j=1$, we can claim that the objective function will be further decreased with all constraints satisfied, which contradicts the fact that the solution is optimal.
If $j>1$,
to ensure constraints (\ref{ener3min1}d) hold for all $l\in\mathcal J_i$, we set\footnote{It is assumed that the maximal transmit power of each user is large enough such that the power of user $j-1$ can be increased} $p_{i(j-1)}'=p_{i(j-1)}^*+\frac{h_{ij}}{h_{i(j-1)}}\epsilon$.
Owing to the fact that $h_{i(j-1)}\geq h_{ij}$, we have $p_{ij}'+p_{i(j-1)}'\leq p_{ij}^*+p_{i(j-1)}^*$.
With new power $p_{ij}'$ and $p_{i(j-1)}'$, the objective function will be further decreased with all constraints satisfied, which contradicts the fact that the solution is optimal.

\vspace{-0.5em}
\section{Proof of Lemma 3}
\setcounter{equation}{0}
\renewcommand{\theequation}{\thesection.\arabic{equation}}
\vspace{-0.5em}

For any optimal solution $(\pmb d^*, \pmb x^*, \pmb \tau^*, \pmb f^*, \pmb p^*, T^*)$ to Problem \eqref{ener3min1} with  $\frac{\tau^*}{x^*}+ \frac{C_{ij}d_{ij}^*}{f_{ij}^*}<T$, we can always construct a new solution $\bar f_{ij}$ with  $\frac{\tau^*}{x^*}+  \frac{C_{ij}d_{ij}^*}{\bar f_{ij}}=T^*$.
We can claim that new solution $(\pmb d^*, \pmb x^*, \pmb \tau^*, \bar{\pmb f}=[f_{11}^*, \cdots, \bar f_{ij}, \cdots, f_{N M_N}^*], \pmb p^*,T^*)$ is feasible with the same objective value of solution $(\pmb d^*, \pmb x^*, \pmb \tau^*, \pmb f^*, \pmb p^*,T^*)$.
Lemma 3 is proved.

\vspace{-0.5em}
\section{Proof of Lemma 4}
\setcounter{equation}{0}
\renewcommand{\theequation}{\thesection.\arabic{equation}}
\vspace{-0.5em}
According to Lemma 3, for the optimal solution to Problem (\ref{ener3min2}), constraints (\ref{ener3min2}b) hold with equality, i.e., $\frac{\tau^*}{x^*}+  \frac{C_{ij}d_{ij} }{\bar f_{ij}}=T^*$,  which yields
\vspace{-0.5em}
\begin{equation}\label{ener3min2eq1}\vspace{-0.5em}
f_{ij}^*=\frac{C_{ij}d_{ij}x_i}{T^* x_i-\tau_i^*}, \quad \forall i\in\mathcal N, j\in\mathcal J_i.
\end{equation}
Considering that $f_{ij}\geq0$ from (\ref{ener3min2}d), we have
$\tau_i^* \leq T^*x_i$,  $\forall i\in\mathcal N$.
Applying \eqref{ener3min2eq1},  Problem  \eqref{ener3min2} becomes:
\vspace{-0.5em}
\begin{subequations}\label{ener3min2eq2}\vspace{-0.5em}
\begin{align}
\mathop{\min}_{ \pmb \tau, T}\;
 \quad&  \omega T + (1-\omega) \sum_{i=1}^N \sum_{j=1}^{M_i}   p_{ij}\tau_i \label{ener3min2eq1a}
  \\
\textrm{s.t.}\quad\qquad \!\!\!\!\!\!\!\!\!
& \sum_{i=1}^N \sum_{j=1}^{M_i}\frac{C_{ij}d_{ij}x_i}{Tx_i-\tau_i} \leq F \label{ener3min2eq1b} \\
&  T_i \leq \tau_i  \leq Tx_i,  T\geq \bar T,  \quad \forall i\in\mathcal N,
\end{align}
\end{subequations}
which can be verified to be a convex problem.
Observing that both objective function \eqref{ener3min2eq1a} and the left term of constraint \eqref{ener3min2eq1b} decreases with $\tau_i$,
the optimal  solution to Problem \eqref{ener3min2eq2} is
\vspace{-0.5em}
\begin{equation}\label{ener3min2eq2_1}\vspace{-0.5em}
\tau_i^*=T_i,  \quad\forall i \in\mathcal N.
\end{equation}
Combining \eqref{ener3min2eq1} and \eqref{ener3min2eq1b}, we have
$T\geq \tilde T$, where $\tilde T$ is the solution to \eqref{ener3min2eq1_2}.
As a result,
the optimal solution of Problem \eqref{ener3min2} is given by \eqref{ener3min2eq1_1}.

\vspace{-0.5em}
\section{Dual Method to Solve Problem \eqref{ener3min5}}
\setcounter{equation}{0}
\renewcommand{\theequation}{\thesection.\arabic{equation}}
\vspace{-0.5em}
The Lagrange function of Problem \eqref{ener3min5} can be given by
\vspace{-0.5em}
\begin{eqnarray}\label{AppGeq1}\vspace{-0.5em}
&&\!\!\!\!\!\!\!\!\!\!\!\!\!\!\!\!
\!\mathcal L_1=
\sum_{i=1}^N\sum_{j=1}^{M_i} (\tau_i p_{ij}  -C_{ij}Q_{ij}d_{ij})
+\sum_{i=1}^N \sum_{j=1}^{M_i}\beta_{ij} \Bigg( \!\frac{\tau_i}{x_i} +\frac{C_{ij}d_{ij}}{f_{ij}}-T
\Bigg)
+\nonumber \\
&&\!\!\!\!\!\!\!\!\!\!\!\!\!\!\!\!
+\sum_{i=1}^N \sum_{j=1}^{M_i} \lambda_{ij} \left( \sum_{l=j}^{M_i} d_{il}- B\tau_i\log_2 \left( 1+\frac { \sum_{l=j}^{M_i} p_{il} h_{il} } {\sigma^2B   } \right) \right)
+ \mu\left(\sum_{i=1}^N x_i - 1
\right)-\sum_{i=1}^N\sum_{j=1}^{M_i}\zeta_{ij}p_{ij}
\nonumber \\
&&\!\!\!\!\!\!\!\!\!\!\!\!\!\!\!\!
+\sum_{i=1}^N\sum_{j=1}^{M_i}\eta_{ij}(p_{ij}-P_{ij})
+\sum_{i=1}^N\sum_{j=1}^{M_i} \theta_{ij}(D_{ij}-d_{ij})+\sum_{i=1}^N \sum_{j=1}^{M_i} \nu_{ij}(d^2_{ij}- R^2_{ij})-\sum_{i=1}^N\rho_i x_i,\nonumber
\end{eqnarray}
where
$\beta_{ij},\lambda_{ij}, \zeta_{ij}, \eta_{ij}, \theta_{ij}, \nu_{ij}, \rho_i \geq  0$ and $\mu$ are Lagrange multipliers associated with the corresponding constraints of Problem \eqref{ener3min5}.
Note that $d_{ij}\leq D_{ij}$ in (\ref{ener3min5}g) is replaced by the equivalent form $d_{ij}^2\leq D_{ij}^2$.
With this replacement, the first-derivative of $d_{ij}$ is not constant, which enables us to obtain a closed-form solution of $d_{ij}$.
To satisfy maximal completion time constraints (\ref{ener3min5}b), we must have $x_i>0$.
Consequently, according to the complementary condition, we can obtain
\vspace{-0.5em}
\begin{equation}\vspace{-0.5em}
\rho_i=0,\quad\forall i \in \mathcal N.
\end{equation}
Based on \cite{boyd2004convex}, the optimal solution should satisfy the following Karush-Kuhn-Tucker conditions of  Problem \eqref{ener3min5}:
\vspace{-0.5em}
\begin{subequations}\label{AppGKKT1}\vspace{-0.5em}
\begin{align}
&
\frac {\partial \mathcal L_1}{\partial d_{ij}} =
-C_{ij}Q_{ij} +\frac{\beta_{ij}C_{ij}}
{f_{ij}}+
\sum_{l=1}^j
\lambda_{il}-\theta_{ij}+2\nu_{ij}d_{ij}=0
\\
&\frac {\partial \mathcal L_1}{\partial x_i}=
- \frac{\sum_{j=1}^{M_i} \beta_{ij}\tau_i}{x_i^2} +\mu=0
 \\
&\frac {\partial \mathcal L_1}{\partial p_{ij}}=
\tau_i- \sum_{l=1}^j
\frac{B\tau_i\lambda_{il} h_{ij}}
{(\ln 2)(\sigma^2 B+\sum_{m=l}^{M_i} p_{im} h_{im}  )}-\zeta_{ij}+\eta_{ij}=0.
\end{align}
\end{subequations}

To solve the convex optimization Problem \eqref{ener3min5}, we use the dual method by iteratively updating the Lagrange multipliers and primary variables.
In the $(n+1)$-th iteration, we can calculate the primary variables with given the Lagrange multipliers $\beta_{ij}(n)$, $\lambda_{ij}(n)$,   $\zeta_{ij}(n)$, $\eta_{ij}(n)$, $\theta_{ij}(n)$, $\nu_{ij}(n)$, and $\mu(n)$.
If $\nu_{ij}(n)>0$,  we can obtain $d_{ij}(n+1)$ based on (\ref{AppGKKT1}a):
\vspace{-0.5em}
\begin{equation}\label{AppGeq2}\vspace{-0.5em}
 d_{ij}(n+1) = \frac{f_{ij}C_{ij}Q_{ij} - {\beta_{ij}(n)C_{ij}}
 -f_{ij}\sum_{l=1}^j\lambda_{il}(n)+f_{ij}\theta_{ij}(n) }
 {2f_{ij}\nu_{ij}(n)}.
\end{equation}
If $\nu_{ij}(n)=0$, according to (\ref{AppGKKT1}a) and (\ref{ener3min5}d), we can obtain
\vspace{-0.5em}
\begin{equation}\label{AppGeq2_2}\vspace{-0.5em}
 d_{ij}(n+1)  = \left\{ \begin{array}{ll}
\!\!\!D_{ij} &  \textrm{if $ \frac {\partial \mathcal L_1}{\partial d_{ij}(n)}  > 0 $}\\
\!\!\!R_{ij} &  \textrm{else}
\end{array} \right. .
\end{equation}
Applying (\ref{AppGKKT1}b), we have
\vspace{-0.5em}
\begin{equation}\label{AppGeq5}\vspace{-0.5em}
x_i(n+1)=\sqrt{\frac{\sum_{j=1}^{M_i}\beta_{ij}(n)\tau_i}{\mu(n)}}, \quad \forall i\in\mathcal N.
\end{equation}
Combining (\ref{sys1min1}e) and \eqref{AppGeq5} yields
\vspace{-0.5em}
\begin{equation}\label{AppGeq5_1}\vspace{-0.5em}
\mu(n)=\left(\sum_{i=1}^N\sqrt{ {\sum_{j=1}^{M_i}\beta_{ij}(n)\tau_i} }\right)^2,
\end{equation}
which shows that Lagrange multiplier $\mu$ can be determined by $\beta_{ij}$.
For the optimal solution to Problem \eqref{ener3min5}, constraints (\ref{ener3min5}c)   hold with equality, as otherwise the objective function can be further improved with all constraints satisfied.
Based on the complementary condition, we can have $\lambda_{ij}(n)>0$.
Solving (\ref{AppGKKT1}c), we can obtain $p_{ij}(n)$ by using the recursion method.

To update the Lagrange multipliers with the primal variables obtained from
\eqref{AppGeq2}-\eqref{AppGeq5} and (\ref{AppGKKT1}c), the gradient based method \cite{bertsekas2009convex} is adopted.
The new values of the Lagrange multipliers are updated by
\vspace{-0.5em}
\begin{eqnarray}\vspace{-0.5em}
&&\!\!\!\!\!\!\!\!\!\!\!\!\!\!\!\!\!\!\!\!\!\!\!\!
\beta_{ij}(n+1)=\left[\beta_{ij}(t)+\delta(n)\!
\left( \!\frac{\tau_i}{x_i(n)} +\frac{C_{ij}d_{ij}(n)}{f_{ij}} -T\!
\right)\right]^+
\\
&&\!\!\!\!\!\!\!\!\!\!\!\!\!\!\!\!\!\!\!\!\!\!\!\!
\lambda_{ij} (n+1)\!=\left[\lambda_{ij} (n)\!+\!\delta(n)  \left( \sum_{l=j}^{M_i} d_{il}(n) - B\tau_i\log_2 \left( 1+\frac { \sum_{l=j}^{M_i} p_{il}(n)  h_{il} } {\sigma^2B   } \right) \right)\right]^+
\\
&&\!\!\!\!\!\!\!\!\!\!\!\!\!\!\!\!\!\!\!\!\!\!\!\!
\zeta_{ij}(n+1)=\left[\zeta_{ij}(n)-\delta(n) p_{ij}(n)\right]^+
\\
&&\!\!\!\!\!\!\!\!\!\!\!\!\!\!\!\!\!\!\!\!\!\!\!\!
\eta_{ij}(n+1)=\left[\eta_{ij}(n)+\delta(n) (p_{ij}(n)-P_{ij})\right]^+
\\
&&\!\!\!\!\!\!\!\!\!\!\!\!\!\!\!\!\!\!\!\!\!\!\!\!
\theta_{ij}(n+1)=\left[\theta_{ij}(n)+\delta(n)(D_{ij}-d_{ij}(n))\right]^+
\\
&&\!\!\!\!\!\!\!\!\!\!\!\!\!\!\!\!\!\!\!\!\!\!\!\!
\nu_{ij}(n+1)=\left[\nu_{ij}(n)+\delta(n)(d^2_{ij}(n)- R^2_{ij})\right]^+,
\end{eqnarray}
where $\delta(t)$ is a dynamically chosen stepsize
and $[x]^+$ denotes $\max\{x,0\}$.
The value of $\mu(n+1)$ is updated according to \eqref{AppGeq5_1}.

\vspace{-0.5em}
\section{Proof of Theorem 1}
\setcounter{equation}{0}
\renewcommand{\theequation}{\thesection.\arabic{equation}}
\vspace{-0.5em}

According to constraints (\ref{time2min1}c), power $p_{ij}$ can be obtained as a function of $x_i t_i$.
To obtain the expression of $p_{ij}$, we set
\vspace{-0.5em}
\begin{equation}\label{AppThe2eq1}\vspace{-0.5em}
 a_{ij}=\sum_{l=j}^{M_i}p_{il} h_{il},\quad \forall k\in\mathcal K.
\end{equation}
Substituting \eqref{AppThe2eq1} into (\ref{time2min1}c), we can obtain:
\vspace{-0.5em}
\begin{equation}\label{AppThe2eq2}\vspace{-0.5em}
 B x_i t_i\log_2\left(
\frac{   a_{ij}+\sigma^2B}
{   a_{i(j+1)} +\sigma^2B}
\right)=D_{ij}.
\end{equation}
According to (\ref{AppThe2eq2}), we have:
\vspace{-0.5em}
\begin{equation}\label{AppThe2eq3}\vspace{-0.5em}
a_{ij}={2^{\frac{D_{ij}}{Bx_it_i}}} a_{i(j+1)}  +\left({2^{\frac{D_{ij}}{Bx_it_i}}} -1\right)\sigma^2B.
\end{equation}

Using the recursive formulation (\ref{AppThe2eq3}) and $a_{i(M_i+1)}=\sum_{l=M_i+1}^{M_i}p_{il}h_{il}=0$, we have:
\vspace{-0.5em}
\begin{equation}\label{AppThe2eq5}\vspace{-0.5em}
a_{ij}= \sum_{l=j}^{M_i} 2^{\frac{\sum_{k=j}^{l-1} D_{ik}}{Bx_it_i}}\left({2^{\frac{D_{il}}{Bx_it_i}}} -1\right)\sigma^2B,
\end{equation}
where we set $\sum_{k=j}^{j-1} D_{ik}=0$.
Based on \eqref{AppThe2eq1}, we have:
\vspace{-0.5em}
\begin{equation}\label{AppThe2eq6}\vspace{-0.5em}
p_{ij}=\frac{a_{ij}-a_{i(j+1)}}{h_{ij}}, \quad\forall i \in \mathcal N, j \in \mathcal J_i.
\end{equation}
Combining \eqref{AppThe2eq5} and \eqref{AppThe2eq6} yields:
\vspace{-0.5em}
\begin{equation}\label{AppThe2eq7}\vspace{-0.5em}
p_{ij}=\frac 1 {h_{ij}}{\left(2^{\frac{D_{ij}}{Bx_it_i}}-1\right)} \sum_{l=j+1}^{M_i} 2^{\frac{\sum_{k=j+1}^{l-1} D_{ik}}{Bx_it_i}}\left({2^{\frac{D_{il}}{Bx_it_i}}} -1\right)\sigma^2B +\frac 1 {h_{ij}}\left({2^{\frac{D_{ij}}{Bx_it_i}}} -1\right)\sigma^2B,
\end{equation}
which monotonically decreases with $x_it_i$.
Considering the maximum uplink transmission power constraints (\ref{time2min1}f),
we can obtain that
\vspace{-0.5em}
\begin{equation}\label{AppThe2eq7_2}\vspace{-0.5em}
x_it_i \geq T_{ij}, \quad\forall i \in \mathcal N, j \in \mathcal J_i.
\end{equation}
where $T_{ij}$ is the solution to equation \eqref{AppThe2eq8}.
With the definition of $\bar T_i$ in \eqref{AppThe2eq9},
\eqref{AppThe2eq7_2} can be further simplified as
\vspace{-0.5em}
\begin{equation}\label{AppAeq3}\vspace{-0.5em}
t_i \geq \frac{\bar T_i}{x_i}, \quad \forall i \in \mathcal N.
\end{equation}
According to constraints (\ref{time2min1}b) and (\ref{AppAeq3}), we can claim that constraints (\ref{AppAeq3}) hold with equality for the optimal solution, i.e.,
$t_i = \frac{\bar T_{i}}{x_i}, \forall i\in\mathcal N$.
Applying (\ref{AppAeq3}) with equality, set (\ref{time2min1}) becomes (\ref{AppBmin2}).
Due to the fact that constraints (\ref{AppBmin2}c)-(\ref{AppBmin2}e) are linear and $  1/ x$ is a convex function, the set \eqref{AppBmin2} is convex.
As a result, Theorem~1 is proved.

\vspace{-0.5em}
\section{Proof of Lemma 5}
\setcounter{equation}{0}
\renewcommand{\theequation}{\thesection.\arabic{equation}}
\vspace{-0.5em}
First, we prove the necessary condition.
Combining (\ref{AppBmin2}b) and (\ref{AppBmin2}e), we have
\vspace{-0.5em}
\begin{equation}\label{AppBeq1}\vspace{-0.5em}
x_{i} \geq \frac{\bar T_i}{T}, f_{ij} \geq \frac {C_{ij}D_{ij}x_i}
{Tx_i-\bar T_i}, \quad \forall i\in\mathcal N, j\in\mathcal J_i.
\end{equation}
Based on (\ref{AppBmin2}c)-(\ref{AppBmin2}e) and (\ref{AppBeq1}), we find that set (\ref{AppBmin2}) is feasible if
\begin{equation}
\sum_{i=1}^N \sum_{j=1}^{M_i} f_{ij} \leq F
\end{equation}
 is satisfied under constraints (\ref{AppBeq1}) and (\ref{AppBmin2}c).
To show this, we can formulate the following minimization problem:
\vspace{-0.5em}
\begin{subequations}\label{AppBmin3}\vspace{-0.5em}
\begin{align}
\mathop{\min}_{\pmb x }\;
 \quad&   \sum_{i=1}^N \sum_{j=1}^{M_i}  \frac {C_{ij}D_{ij}x_i}
{Tx_i-\bar T_i}
  \\
\textrm{s.t.}\quad\qquad \!\!\!\!\!\!\!\!\!
&\sum_{i=1}^N x_i = 1 \\
& x_i \geq \frac{\bar T_i}{T}, \quad \forall i\in\mathcal N.
\end{align}
\end{subequations}
With the help of Problem (\ref{AppBmin3}), we only need to check whether the optimal objective value (\ref{AppBmin3}a) is less than $F$ or not.
To ensure that Problem (\ref{AppBmin3}) is feasible, we must have
\vspace{-0.5em}
\begin{equation}\label{AppBeq1_2}\vspace{-0.5em}
\sum_{i=1}^N  \frac{\bar T_i}{T} \leq 1 .
\end{equation}
Due to the fact that
\vspace{-0.5em}
\begin{equation}\vspace{-0.5em}
\frac{\partial^2 \frac {C_{ij}D_{ij}x_i}
{Tx_i-\bar T_i}}
{\partial x_i^2}
=\frac{2C_{ij}D_{ij}T\bar T_i}
{(Tx_i-\bar T_i)^3}\geq 0, \quad  \forall  x_i \geq \frac{\bar T_i}{T},
\end{equation}
Problem (\ref{AppBmin3}) is a convex problem.
The Lagrange function of Problem (\ref{AppBmin3}) is given by
\vspace{-0.5em}
\begin{equation}\vspace{-0.5em}
\mathcal L_2=\sum_{i=1}^N \sum_{j=1}^{M_i} \frac {C_{ij}D_{ij}x_i}
{Tx_i-\bar T_i} +\alpha \left(\sum_{i=1}^N x_i - 1\right),
\end{equation}
where $\alpha$ is the Lagrange multiplier associated with constraint (\ref{AppBmin3}b).
The first-order derivative of $\mathcal L_2$ can be formulated as
\vspace{-0.5em}
\begin{equation}\vspace{-0.5em}
\frac{\partial \mathcal L_2} {\partial x_i}=  \frac {-\bar T_i\sum_{j=1}^{M_i}C_{ij}D_{ij} }
{(Tx_i-\bar T_i)^2} +\alpha.
\end{equation}
Setting $\frac{\partial \mathcal L_2} {\partial x_i}=0$ yields
\vspace{-0.5em}
\begin{equation}\label{AppBeq2}\vspace{-0.5em}
x_i=  \frac{\bar T_i+\sqrt{\frac {\bar T_i\sum_{j=1}^{M_i}C_{ij}D_{ij} }
{\alpha}} } T .
\end{equation}
Based on (\ref{AppBmin3}b) and \eqref{AppBeq2}, we have
\vspace{-0.5em}
\begin{equation}\label{AppBeq3}\vspace{-0.5em}
\alpha=  \left(
\frac{ \sum_{i=1}^N \sqrt{  {\bar T_i\sum_{j=1}^{M_i}C_{ij}D_{ij} } }}
{T-\sum_{i=1}^N  {\bar T_i} }
\right)^2.
\end{equation}
Combining \eqref{AppBeq2} and \eqref{AppBeq3}, we can calculate the optimal objective value (\ref{AppBmin3}a) as
\vspace{-0.5em}
\begin{align}\label{AppBeq3_2}\vspace{-0.5em}
&\sum_{i=1}^N \sum_{j=1}^{M_i}  \frac {C_{ij}D_{ij}x_i}
{Tx_i-\bar T_i}=\frac{\left( { \sum_{i=1}^N \sqrt{  {\bar T_i\sum_{j=1}^{M_i}C_{ij}D_{ij} } }} \right)^2}
{T\left(T-\sum_{i=1}^N \bar T_i \right)}+\sum_{i=1}^N \sum_{j=1}^{M_i} \frac{C_{ij} D_{ij}}{T}.
\end{align}
Based on \eqref{AppBeq1_2} and \eqref{AppBeq3_2}, the feasibility conditions of Problem (\ref{AppBmin2}) are given by (\ref{time2eq3}) and (\ref{time2eq3_2}).

Next, we prove the sufficient condition.
If conditions (\ref{time2eq3}) and (\ref{time2eq3_2}) are satisfied, we can construct
$\pmb x$ defined in \eqref{AppBeq2} and \eqref{AppBeq3}, and $\pmb f$ defined in \eqref{AppBeq1} with equality. Checking constraints (\ref{AppBmin2}b)-(\ref{AppBmin2}e), we can claim that the constructed solution is feasible, i.e., set \eqref{AppBmin2} is non-empty.

\vspace{-0.5em}
\section{Proof of Theorem 2}
\setcounter{equation}{0}
\renewcommand{\theequation}{\thesection.\arabic{equation}}
\vspace{-0.5em}
We first show the equivalence of Problems \eqref{ener3min1} and \eqref{ener3min1_2}.
When $F=+\infty$, the computation time at the edge $\frac{C_{ij}d_{ij}}{f_{ij}}$ is neglected, i.e., constraints (\ref{ener3min1}c) can be replaced by
\vspace{-0.5em}
\begin{equation}\label{AppHeq1}\vspace{-0.5em}
\tau_i \leq T x_i, \quad\forall i \in \mathcal N.
\end{equation}
Combining \eqref{AppHeq1} and (\ref{ener3min1}e), the equivalent constraint (\ref{ener3min1_2}c) is obtained.
Introducing new variable $q_{ij}=\tau_i p_{ij}$ to replace power $p_{ij}$, $\forall i \in \mathcal N$, $j \in \mathcal J_i$, Problem \eqref{ener3min1} is equivalent to Problem \eqref{ener3min1_2}.

Then, we prove that Problem \eqref{ener3min1_2} is convex.
Obviously, the objective function (\ref{ener3min1_2}a) and constraints (\ref{ener3min1_2}c)-(\ref{ener3min1_2}f) are all linear.
It remains to check the convexity of constraints (\ref{ener3min1_2}b).
Based on \cite[Page~89]{boyd2004convex}, the perspective of $u(\pmb x)$ is the function $v(\pmb x,y)$ defined by
\vspace{-0.5em}
\begin{equation}\vspace{-0.5em}
 v(\pmb x,y)=yu(\pmb x/y), {\textbf{dom}}\:v= \{(\pmb x,y)|\pmb x/y\in {\textbf{dom}}\:u, y>0\}.
 \end{equation}
If $u(\pmb x)$ is a concave function, then so is its perspective function $v(\pmb x,y)$ \cite[Page~89]{boyd2004convex}.
Define function
\begin{equation}
u_{ij}(q_{ij}, \cdots, q_{iM_i} )=B \log_2 \left( 1+\frac { \sum_{l=j}^{M_i} q_{il} h_{il} } {\sigma^2B   } \right), \quad \forall i \in \mathcal N, j \in \mathcal J_i.
\end{equation}
Since function $u_{ij}(q_{ij}, \cdots, q_{iM_i})$ is concave w.r.t. $(q_{ij}, \cdots, q_{iM_i})$, the perspective function
\vspace{-0.5em}
\begin{equation}\vspace{-0.5em}
\tau_i
u_{ij} \left(\frac{q_{ij}}{\tau_i},\cdots,  \frac{q_{iM_i}} {\tau_i}\right)
=B \tau_i\log_2 \left( 1+\frac { \sum_{l=j}^{M_i} q_{il} h_{il} } {\sigma^2B  \tau_i } \right)
\end{equation}
is concave w.r.t.
$(q_{ij}, \cdots, q_{iM_i},\tau_i)$.
As a result, constraints (\ref{ener3min1_2}b) are convex.

\vspace{-0.5em}
\section{Proof of Lemma \ref{maxtime}}
\setcounter{equation}{0}
\renewcommand{\theequation}{\thesection.\arabic{equation}}
\vspace{-0.5em}

We first define function $y=x \ln\left( 1+ \frac{1}{x}
\right), x>0$.
Then, we have
\vspace{-0.5em}
\begin{equation}\label{AppEeq2}\vspace{-0.5em}
 y'=\ln\left( 1+ \frac{1}{x}
\right) -\frac{1}{x+1},
 y''= -\frac{1}{x(x+1)^2}<0.
\end{equation}
According to  \eqref{AppEeq2}, $y'$ is a  decreasing function.
Combining $\lim_{t_i \rightarrow +\infty} y' =0$ from (\ref{AppEeq2}) and $y'$ is a  decreasing function, we can obtain that $y'>0$ for all $0<x<+\infty$.
As a result, $y$ is an increasing function.

We then prove that $\sum_{i=1}^N t_i^*=T$ for the optimal solution to problem (\ref{ener3min1}) by using the contradiction method.
Suppose that the optimal solution $(\pmb d^*, \pmb \tau^*, \pmb q^*, T^*)$ to Problem (\ref{ener3min1_2}) satisfies $\sum_{i=1}^N \tau_i^*<T^*$.
We can increase $\tau_1^*$ to $\bar \tau_1=\tau_1^*+\kappa_1$ $(\kappa_1>0)$ such that $\bar \tau_1 + \sum_{i=2}^N \tau_i^*= T^*$.
Since function $y$ is an increasing function, we can slightly decreases $ q_{11}^*$   to  $\bar q_{11}= q_{11}^*-\kappa_2$ $(\kappa_2>0)$ such that
\vspace{-0.5em}
\begin{subequations}\label{AppEeq3}\vspace{-0.5em}
\begin{align}
&B\bar \tau_1 \log_2 \left( 1+\frac { \bar q_{11}h_{11} + \sum_{l=2}^{M_1} \bar q_{1l}^* h_{1l} } {\sigma^2B \bar \tau_1   } \right)
=B \tau_1^* \log_2 \left( 1+\frac {  \sum_{l=1}^{M_1} \bar q_{1l}^* h_{1l} } {\sigma^2B  \tau_1^*   } \right)
\geq \sum_{l=1}^{M_1} d_{1l} ^* .
\end{align}
\end{subequations}
With new solution
$
(\pmb d^*\!, \bar {\pmb \tau}\!=\![\bar \tau_1, \tau_2^*,\cdots, \tau_N^*] , \bar {\pmb q}\!=\![\bar q_{11}, q_{12}^*, \cdots, q_{NM_{N}}^*], T^*)$,
 we can claim that the new solution is feasible with lower objective value, which contradicts that $(\pmb d^*, \pmb \tau^*, \pmb q^*, T^*)$ is the optimal solution to Problem (\ref{ener3min1_2}).

As a result, for the optimal solution to Problem (\ref{ener3min1_2}), we have $\sum_{i=1}^N \tau_i^*=T^*$.


\bibliographystyle{IEEEtran}
\bibliography{IEEEabrv,MMM}
\end{document}